\documentclass[11pt]{article}
\usepackage{moriond,epsfig}
\usepackage{amsmath,amssymb,amsfonts,amsbsy}
\usepackage[square,comma,numbers]{natbib}
\usepackage{subfigure}

\def\be{\begin{equation}}
\def\ee{\end{equation}}
\def\bea{\begin{eqnarray}}
\def\eea{\end{eqnarray}}

\begin{document}
\vspace*{4cm}
\title{GAW - A very large field-of-view Imaging Atmospheric Cherenkov Telescope}
\vspace{-0.2cm}
\author{ Lu\'{\i}sa Arruda, \small{on behalf of the GAW collaboration}}

\address{LIP \\
         Av. Elias Garcia, 14, 1$^o$ andar\\
         1000-149 Lisboa, Portugal \\
         e-mail: luisa@lip.pt}

\maketitle\abstracts{
GAW (Gamma Air Watch) is a pathfinder experiment in the TeV range to test
the feasibility of a new generation of Imaging Atmospheric Cherenkov
Telescopes (IACT). It combines high flux sensitivity with large field-of-view
(FoV =
24$^{\textnormal{\footnotesize{o}}}$$\times$24$^{\textnormal{\footnotesize{o}}}$
) using Fresnel lenses, stereoscopic observational approach and single-photon counting mode. This particular counting
mode, in comparison with the usual charge integration one, allows the
triggering of events with a smaller number of collected Cherenkov
photons keeping a good signal/background separation. GAW is conceived as an array of three identical imaging telescopes
with 2.13\,m diameter placed at the vertices of an equilateral
triangle of 80\,m side.  The telescope will be built at the Calar Alto
Observatory site (Sierra de Los Filabres - Almeria Spain, 2168\,m a.s.l.) and is a joint effort of research
institutes in Italy, Portugal and Spain. The main characteristics of the
experiment will be reported.
}

\section{Introduction}
High energy gamma-rays are a powerful probe for several astrophysical
quests. During the recent years a new window has been opened in the
observation of gamma-rays from about few tens of MeV up to EeV thanks to the
availability of new photon detectors built using technologies imported from
experimental particle physics. Satellites cover the lowest energy range of
detection (few MeV up to few tens GeV) while ground-based detectors like Imaging Atmospheric Cherenkov
Telescopes (IACTs) and Extensive Air Shower (EAS) arrays cover
higher energy regions (the former from 50\,GeV up to more than 10\,TeV and
the latters with a threshold at 0.5-1\,TeV). The aim of this article is
to introduce a Research and Development experiment for the construction of a
new IACT to detect very high energy (VHE) gamma-rays \footnote{As a rule of
  thumb, VHE $\gamma$-rays are classified as $\gamma$-rays with
energies from $\sim$30\,GeV up to $\sim$30\,TeV \cite{bib:deangelis-2008}}. Although this is a young research field, VHE gamma-ray astronomy
is a well established discipline with several identified sources, steady and
variable, galactic and extragalactic. 

The existing and planned ground-based
IACT observatories aim to lower the energy threshold to few tenths of GeV to
overlap with satellite detection region. The second goal is to improve the
flux sensitivity in the region above 100\,GeV through a stereoscopic
observational approach. Another important purpose is to do a full sky
coverage since astronomical events can occur at unkown locations and/or
randomly in time. A current IACT telescope consists of an optical system with 
few degrees field-of-view (FoV $\le 5^{\textnormal{\footnotesize{o}}}$) and
of a pixelized camera placed at its focus. They can not achieve larger FoV 
due to mirror optical aberrations that rapidly increase with off-axis
angles. Moreover, the increasing of the detector area required to cover large
FoV would unavoidably produce a strong reduction of the light collecting area
due to the shadow of the detection matrix onto the reflector. Such
limitation has to be overcomed to significantly improve the capability of
surveying large sky areas since detection of transient phenomena is a goal. 
An alternative solution might be the usage of refractive optics like
Fresnel lenses as light collectors instead of the classical mirror. Fresnel
lenses enable large FoV, they have good transmittance and avoid the shadow
problem. Since they are easy to replicate Fresnel lenses appear as an
affordable solution however chromaticity should be controlled.
The study of the feasibility of such solution is the aim of the GAW experiment as explained below.   

\subsection{The GAW experiment}\label{subsec:GAWexperiment}
\begin{figure}[ht]
\vspace{-0.5cm}
\begin{minipage}[b]{0.5\linewidth}
\centering
\includegraphics[width=0.8\textwidth,angle=-90,bb=14 14 444 439]{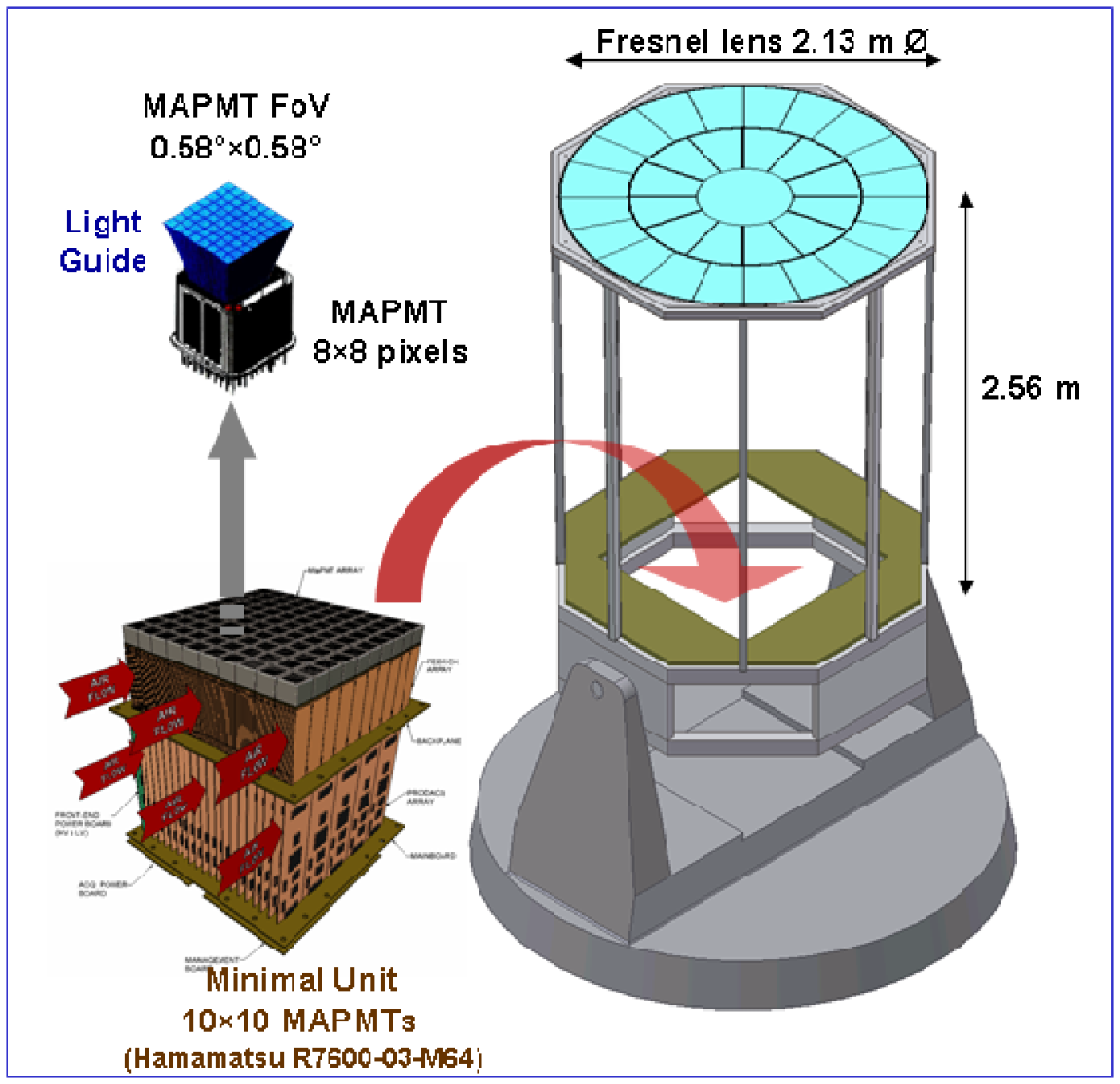}
\caption{Schematic view of the GAW telescope.}
\label{fig:GAWtelescope}
\end{minipage}
\hfill
\hspace{0.5cm}
\begin{minipage}[b]{0.5\linewidth}
\centering
\hspace{1.9cm}
\vspace{0.2cm}
\scalebox{0.45}{\includegraphics{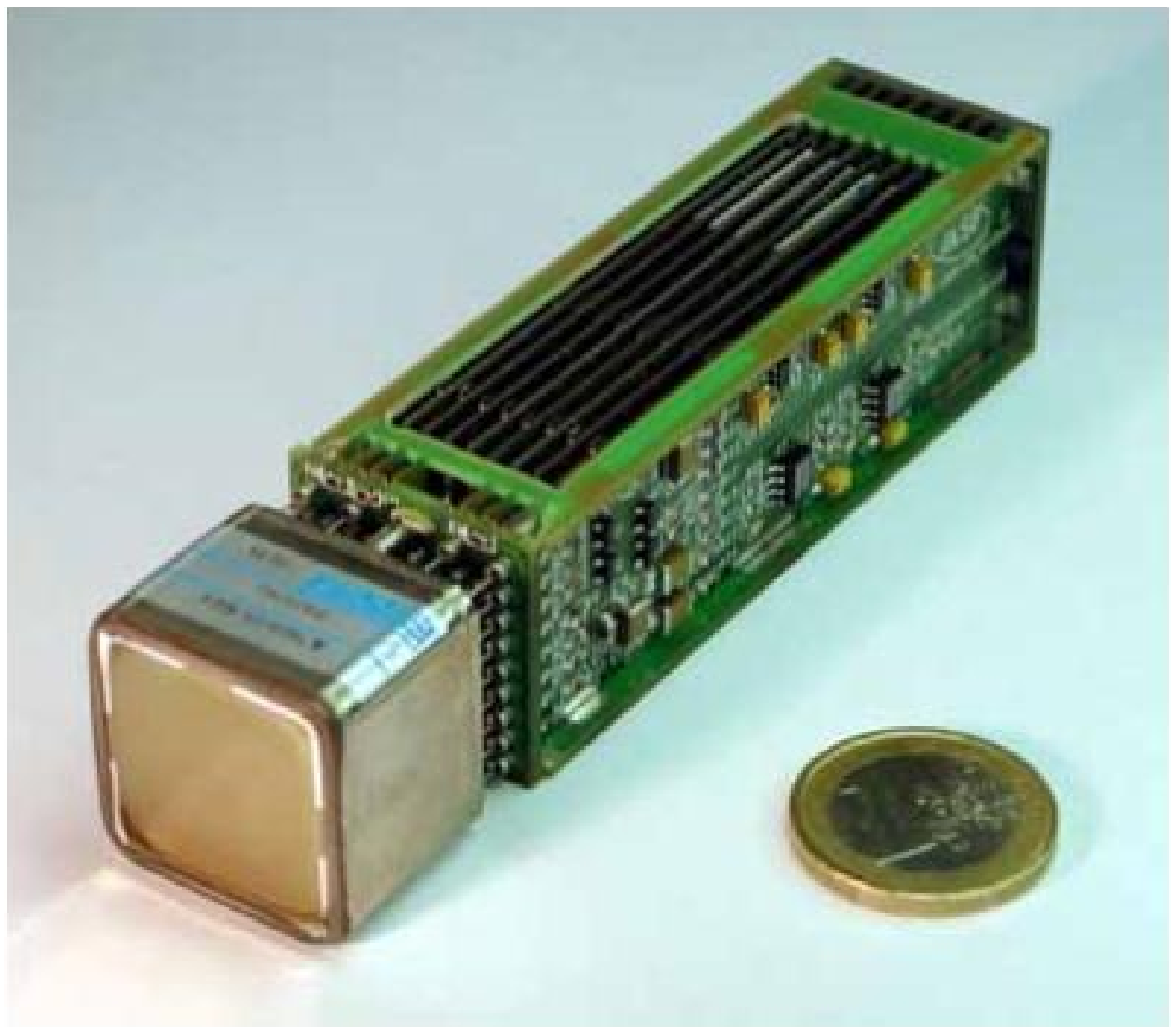}}
\vspace{-0.2cm}
\caption{FEBrick with a MultiAnode PMT}
\label{fig:MAPMT}
\end{minipage}
\hfill
\end{figure}

The design of the GAW telescope includes a Fresnel lens and a focal surface
detector formed by a grid of {\bf M}ulti{\bf A}node pixelized (8x8)
{\bf P}hoto{\bf M}ultiplier
{\bf T}ubes (MAPMT) coupled to light guides. A schematic view of the GAW telescope
is depicted in \mbox{Figure \ref{fig:GAWtelescope}}. A detailed description
of the GAW detector is given in reference \cite{bib:Proposal}. 
The design of the GAW experiment was made to prove the feasibility of the usage of a Fresnel lense as
an efficient light collector allowing an enlargement of the field of view.
Another innovative idea is that instead of the usual charge integration
method, GAW front-end electronics design will be based on single photoelectron
counting mode \cite{bib:speOCatalano}. In such working mode, the effects of the electronic noise and
the photomultiplier gain differences are kept negligible. This method strongly reduces the
minimum number of photoelectrons (p.e.) required to trigger the system and,
consequently a low telescope energy threshold ($\sim 700$\,GeV) is achieved despite
the relatively small dimension of the Cherenkov light-collector (2.13\,m diameter). The 
pixel size is small enough (3.3425\,mm) to reduce the photoelectrons pile up 
within intervals shorter than sampling time (10\,ns). Current camera design is confortable with
a threshold of 14 p.e per event per trigger-cell (2$\times$2 multianode photomultipliers) since the expected night
sky background (NSB) contribution is 2-3 p.e. per sample per trigger-cell.

The light collector is a
non-commercial Fresnel lens with focal
length of 2.56\,m and 3.2\,mm thick. The lens is made of UltraViolet
transmitting polymetacrilate with a nominal transmitance of $\sim 95\%$ from 330\,nm
to 600\,nm. The lens design is optimized for the maximum wavelength of photon
detection ($\lambda\sim360$\,nm). The lens is composed of a central core
($\varnothing$ 50.8\,cm) surrounded by a corona of 12 petals extending for
40.6\,cm and by a second level of 20 petals for the outer corona extending for
more 40.6\,cm. A mechanical spider support will keep all pieces together.

The MAPMT used for GAW focal surface detector is the Hamamatsu R7600-03-M64
(\mbox{Figure \ref{fig:MAPMT}}) 
with 64 anodes arranged in an $8\times8$ matrix. The physical dimension of
the tube section is $25.7\times25.7$\,mm$^2$ while the effective area is 
$18.1\times18.1$\,mm$^2$. The
tube is equipped with a bialkali photocathode and a 0.8\,mm thick UV-transmitting window (from 200 up to 680\,nm). This ensures good quantum
efficiency for wavelengths longer than 300\,nm with a peak of 20\% at
420\,nm. The Metal Channel Dynode structure with 12 stages provides a gain of
the order of $3\times10^5$ for an applied voltage of 800\,V. This PMT provides a
fast response (of the order of 10\,ns) in order to disentangle the Cherenkov
light, which is produced coherently in space and time, from the incoherent but
significantly fluctuating NSB.

In order to reduce dead areas between adjacent photomultipliers and
consequently to increase the photon collection efficiency, an array of light
guides was added, coupled to each photomultiplier. Due to the dead areas
between adjacent PMTs around 55\% of the photons
would be lost without any guiding device. A light guide unit is a
pyramidal polyhedron composed of $8\times8$ independent, plastic tubes glued on a
plastic plate. The tubes are made of a polymetacrilate (PMMA) from Fresnel
Technologies with a refractive index of 1.489 close to the one of
the PMT window ($n=1.5$). These characteristics were chosen to obtain a
transmittance as high as possible over the wavelength range of the PMT
detection. The 64 pieces that constitute the light guide, with ten different shapes, are held
together by a thin layer (1\,mm) on the top made of an anti-reflective PMMA. 
Inside the light guide, photons are conducted by internal reflections. The
light guide unit is optically coupled to the active area of phototube cathode
through a 1\,mm flexible optical pad. With a total height of 35\,mm, and a
collecting surface of 26.74$\times$26.74\,mm$^2$, it presents a readout pixel
size of 3.3425\,mm which renders in a spatial granularity of
$\sim0.1^{\textnormal{\footnotesize{o}}}$ suitable for Cherenkov imaging. The optimum dimensions have been determined to maximize
the photon collection efficiency ($\sim$71\%), to minimize the cross talk between adjacent
pyramidal frustuns ($\sim$6.5\%) and to achieve the higher spatial uniformity in the photon
collection efficiency (uniform at the level of 0.01). All these parameters
were evaluated with simulated samples for the photon
incident angles on the top of the light guide within the GAW FoV.

\subsection{The GAW project timeline}

\begin{figure}[ht]
\vspace{-1.0cm}
\begin{minipage}[b]{0.45\linewidth}
\centering
\vspace{-2.8cm}
 \subfigure[]{\label{fig:GAWmechanics}\includegraphics[width=1.05\textwidth,angle=0,bb=14 14 404 412]{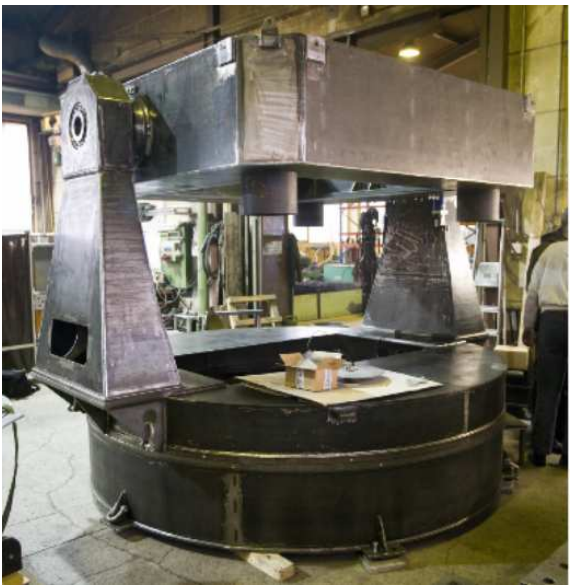}}
\vspace{1.2cm}
\end{minipage}
\hfill
\hspace{0.5cm}
\begin{minipage}[b]{0.45\linewidth}
\centering
\vspace{-0.4cm}
\subfigure[]{\label{fig:Housing}\scalebox{0.25}{\includegraphics{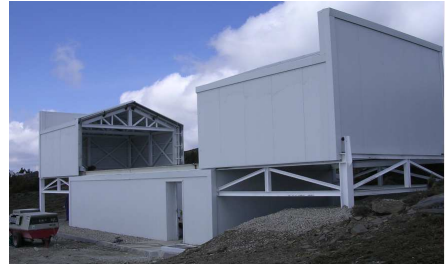}}}
\vspace{0.2cm}
\centering
\vspace{-0.4cm}
\subfigure[]{\label{fig:RackFrame}\scalebox{0.2}{\includegraphics{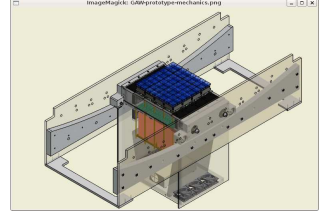}}}
\vspace{1.2cm}
\end{minipage}
\hfill
\vspace{-1.8cm}
\caption{(a) GAW telescope mechanical structure. (b) Telescope housing in
  Calar Alto. (c) Artistic view of the detector focal surface mounted on a rack frame.}
\label{richdet}

\end{figure}

The chosen site for the telescope placement is the Calar Alto Observatory
(Sierra de Los Filabres, Almeria, Spain), at 2168\,m above sea level. The
civil engineeering work at the Calar Alto site is close to be finished and,
in particular, the construction of the building to house the telescope is
finished as can be seen in \mbox{Figure \ref{fig:Housing}}. The
telescope mechanical structure, which is depicted in \mbox{Figure
  \ref{fig:GAWmechanics}}, and the spider support for the lens petals were
manufactured by a specialized company (ASTELCO Systems) and shipped to the 
site, after validation tests carried at the company headquaters. The telescope and lens comissioning will be undertaken
in 2009/2010. The main goal for this stage is to prove the feasibility of the
GAW concept, in particular the optics and the data acquisition systems. A
reduced Fresnel lens with a single central petal  will be installed in the centre of
the supporting spider and the lens design will be validated by measuring the
spot size with Vega spectrum. The first tests of GAW electronics will also be carried
out.
 Afterwards the project will start with a first phase where only one telescope will be
 assembled, with a reduced focal surface detector (10$\times$10 MAPMT),
 covering a FoV of 6$^{\textnormal{\footnotesize{o}}}$$\times$6$^{\textnormal{\footnotesize{o}}}$. This phase,
 starting in 2010, will be suited to test this detector principle in
 ``on-axis'' mode and in ``off-axis'' observation mode. The focal
 surface will be mounted on a rack frame (\mbox{Figure
 \ref{fig:RackFrame}}) and moved to enable sensitivity measurement by observing the Crab Nebula, on-axis and
 off-axis, up to 12$^{\textnormal{\footnotesize{o}}}$, the edge of the GAW FoV. 
Obtained the R$\&$D results and in case of a sucessful confirmation of the
 GAW concept a second phase is foreseen, with a fully equipped focal plane with
 24$^{\textnormal{\footnotesize{o}}}$$\times$24$^{\textnormal{\footnotesize{o}}}$ FoV. For this phase three identical
 telescopes will be constructed, placed at the vertexes of a
 equilateral triangle (80\,cm side) and will work in the stereoscopic
 mode, improving the angular resolution, the cability of identifying
 gamma-ray induced showers and the determination of the primary photon energy.

\section{Summary}
IACTs with large FoV will offer two important advantages: they will survey the sky for serendipitous
TeV detections and, at the same time, will increase the IACT collection area, triggering events whose
core is far away from the telescope axis and therefore improving the statistics of the high energy tail
of the source spectra. Presently, GAW is a R\&D experiment to build a Cerenkov
telescope that will test the feasibility of a new generation IACT that joins large FoV and
high flux sensitivity. Large FoV will be achieved by using refractive optics made of
Fresnel lens of moderate size. The focal camera will use
the single photon counting mode instead of the charge integration mode widely used in the present IACT experiments. This
working mode will allow the detector to be operated with a low photoelectron
threshold and a consequent lowering of the energy threshold. The stereoscopic observational approach will improve
the angular resolution. GAW is a collaboration effort of several Research
Institutes in Italy, Portugal and Spain. It will be erected in the Calar Alto
Observatory (Sierra de Los Filabres - Andalucia, Spain). The first telescope
is forseen for 2010.

\section{Acknowledgments}
I would like to express my acknowledgments to the GAW collaboration for
giving me the opportunity to attend the 44$^{\textnormal{th}}$ Rencontres de Moriond on {\em Very High Energy Phenomena in the Universe}. I also
express my gratitude to the organizers for the financial support to participate in the meeting.


\end{document}